\documentclass[secnumarabic,amssymb, nobibnotes, aps, prd, preprint]{revtex4-1}

\usepackage[version=3]{mhchem} 
\usepackage{graphicx}
\usepackage{dcolumn}
\usepackage{bm}
\usepackage{upgreek}

\begin{document}

\title{High-Speed Electro-Optic Modulator Integrated with Graphene-Boron Nitride Heterostructure and Photonic Crystal Nanocavity}

\author{Yuanda Gao}
\affiliation{Department of Mechanical Engineering, Columbia University}
\altaffiliation{These authors contribute equally to this work}

\author{Ren-Jye Shiue}
\affiliation{Department of Electrical Engineering and Computer Science, Massachusetts Institute of Technology}
\altaffiliation{These authors contribute equally to this work}

\author{Xuetao Gan}
\affiliation{Department of Electrical Engineering, Columbia University }

\author{Luozhou Li}
\affiliation{Department of Electrical Engineering and Computer Science, Massachusetts Institute of Technology }

\author{Cheng Peng}
\affiliation{Department of Electrical Engineering and Computer Science, Massachusetts Institute of Technology }

\author{Inanc Meric}
\affiliation{ Department of Electrical Engineering, Columbia University }

\author{Lei Wang}
\affiliation{Department of Mechanical Engineering, Columbia University }

\author{Attila Szep}
\affiliation{Air Force Research Laboratory, Sensors Directorate, WPAFB}

\author{Dennis Walker, Jr.}
\affiliation{Air Force Research Laboratory, Sensors Directorate, WPAFB}

\author{James Hone}
\affiliation{ Department of Mechanical Engineering, Columbia University }

\author{Dirk Englund}
\affiliation{ Department of Electrical Engineering and Computer Science, Massachusetts Institute of Technology }
\email{englund@mit.edu}

\date{Dec 17, 2014}%

\begin{abstract}

Nanoscale and power-efficient electro-optic (EO) modulators are essential components for optical interconnects that are beginning to replace electrical wiring for intra- and inter-chip communications\cite{Jalali2006,Lee2010,Reed2010,2009.Miller}. Silicon-based EO modulators show sufficient figures of merits regarding device footprint, speed, power consumption and modulation depth\cite{Liu2004a,Xu2005b,Jacobsen2006,Xu2007a,Alloatti2011,Dong2012,Reed}. However, the weak electro-optic effect of silicon still sets a technical bottleneck for these devices, motivating the development of modulators based on new materials. Graphene, a two-dimensional carbon allotrope, has emerged as an alternative active material for optoelectronic applications owing to its exceptional optical and electronic properties\cite{Bonaccorso2010,Avouris2014a,Bao2012}. Here, we demonstrate a high-speed graphene electro-optic modulator based on a graphene-boron nitride (BN) heterostructure integrated with a silicon photonic crystal nanocavity. Strongly enhanced light-matter interaction of graphene in a submicron cavity enables efficient electrical tuning of the cavity reflection. We observe a modulation depth of 3.2 dB and a cut-off frequency of 1.2 GHz.

\end{abstract}

\maketitle

As an active optical material, graphene exhibits highly desirable uniform absorption over a broad spectral range from visible to mid-infrared\cite{Kuzmenko2008,Mak2008b}, which can be dramatically suppressed by electrostatic doping, enabling electro-absorptive modulation of the incident light\cite{Wang2008}. In addition, the remarkably high carrier mobility in graphene promises high-speed operation and low power consumption\cite{Dean2010b,Xia2009a}. To date, several prototype graphene-based EO modulators have been realized by integrating graphene with silicon waveguides or cavities\cite{Liu2011d,Liu2012m,Gan2013a,Majumdar2012a}. The waveguide-integrated modulators commonly require a long active graphene channel of 25 to 40 $\upmu$m\cite{Liu2012m,Liu2011d}. Therefore, the power consumption is intrinsically high due to the large device footprint and capacitance. For cavity-integrated modulators, graphene strongly interacts with the cavity resonant field, resulting in strong modulation of the cavity reflection with only a few micrometers device footprints\cite{Gan2013a,Majumdar2012a}. However, these devices require electrolyte doping to change the graphene Fermi energy, thus the modulation speed has been limited to several kHz.

Here, we demonstrate a modulator architechture based on a high-mobility dual-layer graphene capacitor integrated with a planar photonic crystal (PPC) cavity. The PPC cavity greatly amplifies the absorption of light into the two graphene sheets. The modulation occurs as the top and bottom graphene layers are oppositely doped by the induced electrostatic potential; when the respective Fermi levels are shifted to half of the incident photon energy ($\hbar\omega/2$), Pauli blocking suppresses the optical absorption\cite{Li2008b,Mak2008b}. We achieved a maximum modulation depth of 3.2 dB within a voltage swing of only 2.5 V. Based on the alternation of cavity's resonance and absorption\cite{Gan2012}, we deduced the optical conductivity of the graphene capacitor as a function of the applied gate voltage and found good agreement with theoretical predictions. The modulation speed is greatly enhanced over previously demonstrated cavity-integrated graphene modulators relying on electrolyte doping; we measured a 3dB cut-off frequency up to 1.2 GHz, which appears to be dominated by the RC time constant of the device. 

\begin{figure}
  \includegraphics[width=17.0cm]{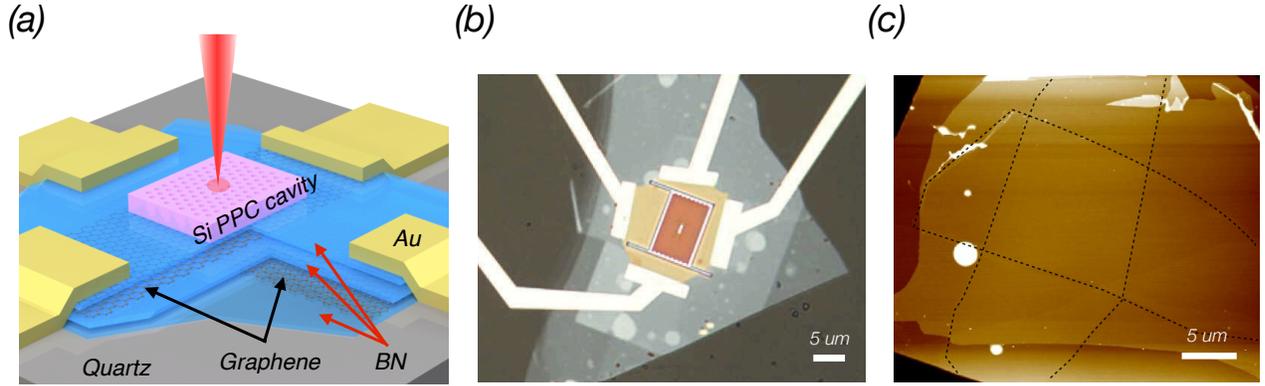}
  \caption{(a) Schematic of the cavity-graphene electro-optic modulator. The dual-layer graphene capacitor on quartz substrate is optically coupled to the PPC cavity. (b) Optical image of fabricated cavity-graphene electro-optic modulator. (c) AFM image of the BN-graphene-BN-graphene-BN five-layer stack. Two graphene layers (black dashed line) are pristine and sandwitched in the BN layers.}
\end{figure}

Fig. 1(a) shows the schematic of the cavity-integrated graphene electro-optic modulator. A BN/Graphene/BN/Graphene/BN five-layer stack was built by the van der Waals (vdW) assembly technique\cite{Wang2013b} and then transferred onto a quartz substrate, see Fig. 1(c). Quartz substrates reduce parasitic capacitance compared to more commonly used  SiO$_2$/Si substrates. The two graphene sheets were positioned as crossed stripes in order to be contacted individually. The graphene edges were exposed by plasma etching\cite{Wang2013b} the five-layer stack using a hydrogen-silsesquioxane (HSQ) resist mask patterned by electron beam lithography (EBL). Metal contacts of Cr/Pd/Au (1/20/50 nm) were deposited by electron beam evaporation, making edge-contact to the two graphene sheets. In this encapsulated dual-layer graphene structure, each one of the graphene sheets can be viewed as a gate and supplys gate voltage to another. The graphene-BN heterostructures made by the vdW technique have demonstrated extremely high room temperature mobility up to 140,000 cm$^2$/Vs, which is comparable to the theoretical acoustic phonon-scattering limit\cite{Wang2013b}.

The PPC cavity was separately fabricated on a silicon-on-insulator (SOI) wafer using a combination of EBL, dry etching, and wet etching. The membrane of the PPC cavity has a thickness of 220 nm and a lattice period of $a$ = 420 nm and air hole radius $r$ = 0.29 $a$. A linear three-hole (L3) defect defines the cavity\cite{Akahane2003}. The PPC lattice was surrounded by a 1-$\upmu$m-wide trench and only connected to the plane of the wafer by four 200-nm-wide bridges, which allowed us to separate them from the parent wafer using an adhesive, transparent Polydimethylsiloxane (PDMS) substrate. Then, using the same vdW technique, we transferred the PPC cavity onto the top surface of the pre-fabricated five-layer-stack device. The PPC cavity was aligned with the overlapping area of the cross-placed two graphene sheets underneath. The completed modulator is shown in Fig. 1(b).

We characterized the PPC cavity using a cross-polarized confocal microscope with a broad-band (super-continuum laser) excitation source, as shown in Fig. 2(a). Before transfer, the suspended cavity membrane exhibited two prominent modes at 1519 nm and 1538 nm with quality factors ($Q$) of 1500 and 1100, respectively, see blue curve in Fig. 2(c). The polarization of the far-field radiation of the resonance at 1538 nm is orthogonal to the long axis of the L3 cavity, which, together with its energy, indicates the fundamental mode\cite{Chalcraft2007}. The energy density of this mode, obtained by finite-difference time-domain (FDTD) simulation, is shown in Fig. 2(b). After the cavity was picked up by the PDMS substrate, the cavity resonance red-shifted, and the $Q$ values dropped to 1050 and 1020, respectively, as seen in the green curve in Fig. 2(c). The red-shift of resonances and decrease of the $Q$ values are expected due to the increase of the effective cavity mode index and the critical angle for total internal reflection. After the cavity was finally transferred onto the graphene capacitor, shown as red curve in Fig. 2(c), the cavity mode exhibited additional red-shift because of the larger refractive index of quartz ($n=1.53$) compared to PDMS ($n=1.4$ at 1550~nm), and the $Q$ values sharply descreased to 270 and 250, primarily due to the graphene absorption.

\begin{figure}
  \includegraphics[width=17.0cm]{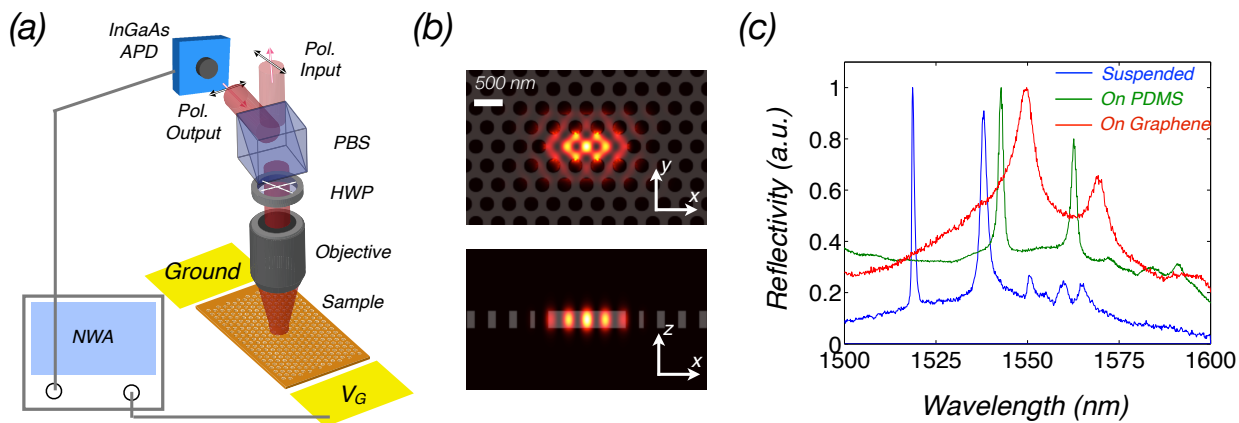}
  \caption{(a) A schematic of the cross-polarized confocal microscope measurement setup. (b) A Finite-difference time-domain (FDTD) simulation of the cavity energy intensity. (c) Reflection spectra of the PPC cavity suspended over SOI substrate (blue), on PDMS substrate (green) and on the dual-layer graphene capacitor (red).}
\end{figure}

We tested the optical response of the graphene modulator by monitoring the cavity reflection spectrum while slowly sweeping the gate voltage $V_G$ across the top and bottom graphene layer, as shown in Fig. 3(a). When $V_G$ increased from 2.5 V to 6.7 V, the carrier density in the two graphene layers gradually increased, approaching the Fermi energy ($\hbar\omega/2$) of Pauli blocking for an incident photon with angular frequency $\omega$, as depicted in Fig. 3(b). Therefore, the optical absorption in graphene was reduced, resulting in increased reflection from the cavity and slightly red-shifted peaks. At $V_G$ = 6.7 V, the cavity reflectivity was increased by 2.1 times compared to the near zero-bias regime, i.e., 0 V $< V_G <$ 2.5 V, corresponding to a maximum modulation depth of 3.2 dB . The spectra for negative bias voltage, -5.6 V $< V_G <$ 0 V, mimicked the spectra for positive voltage; note that for $V_G <$ 0 V, the doping types of top and bottom graphene layers were reversed compared to $V_G >$ 0 V. We plot the cavity reflection at $\lambda = 1551$ nm as a function of $V_G$ in Fig. 3(c). The evolution of the spectrum is symmetric to $V_G =$ 0.2 V, indicating low intrinsic doping of the encapsulated dual-layer graphene capacitor.

By fitting the spectrum in Fig. 3(a) to Lorentzian curves, we obtained the variation of $Q$ values and shift of resonant wavelength as a function of V$_G$. The results are plotted in Fig. S1(a) and Fig. S1(b). The variation of $Q$ values and shift of resonance wavelength are related to the gate-dependent complex optical conductivity $\sigma_g$ of the graphene sheets. We extracted the real $\sigma_{gr}$ (green) and imaginary $\sigma_{gi}$ (black) components of $\sigma_g$ from the cavity spectra using the perturbation theory model\cite{Gan2012,Gan2013a}, as seen in Fig. S1(c) (see supporting information). We notice that the maximum gate voltage applied to the graphene sheets only reduced $\sigma_{gr}$ to half of the value of the undoped graphene, i.e., when $V_G=0$. This indicates that the modulation depth could likely be further increased with a larger gate voltage. However, to avoid breakdown ($V_{breakdown}$ = 0.3 $\sim$ 0.8 V/nm\cite{Lee2011f,Voskoboynikov1976}) of the 10 nm central BN layer, we did not push the gate voltage further more. The limitation of the breakdown voltage could be improved by working at longer wavelengths or with high-K materials that provide stronger electric field between two graphene sheets before breakdown.

\begin{figure}
  \includegraphics[width=17.0cm]{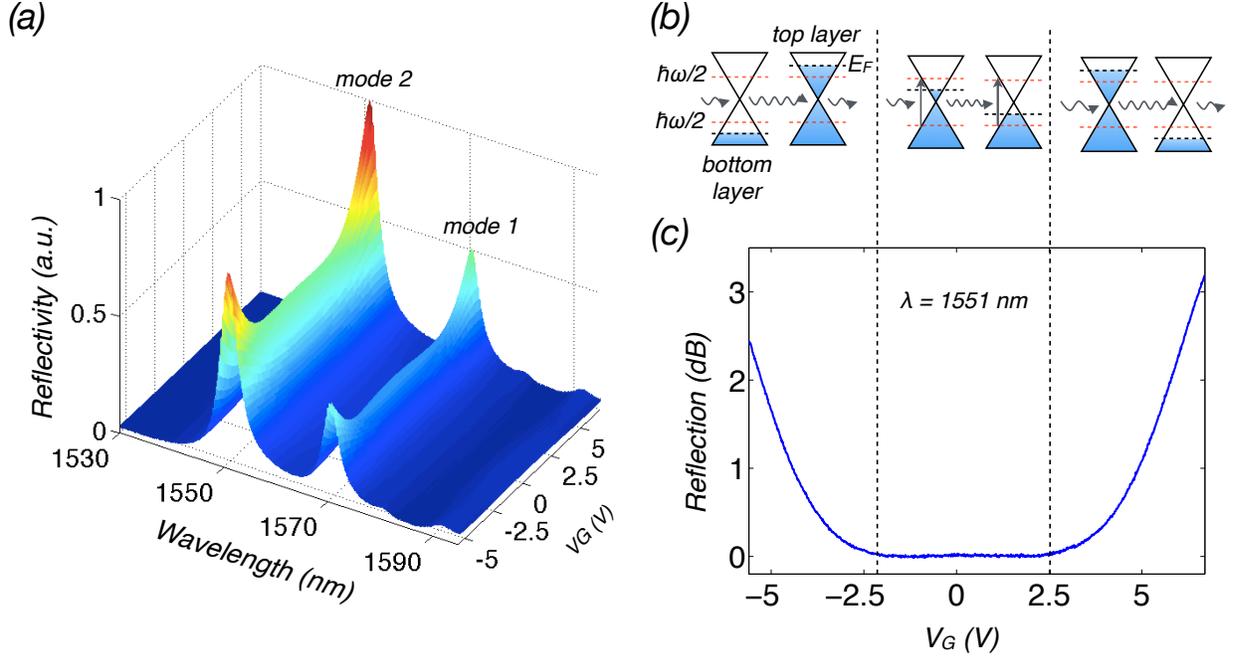}
  \caption{(a) Cavity reflection spectrum as a function of gate voltage $V_G$ and wavelength $\lambda$. Two peaks at 1551 nm and 1570 nm show significant intensity increase when the gate voltage $|V_G|$ increases. (b) Band diagram of two graphene sheets at different gate voltage. In the region where $V_G>$ 2.5 V or $V_G<$ -2.1 V, the Fermi level ($E_F$, black dashed line) is either lower or higher than the half photon energy ($\hbar\omega/2$, red dashed line) for bottom or top graphene layer, thus suppressing the probability of interband transition and making graphene more transparent to the incident photon. In the region where -2.1 V $<V_G<$ 2.5 V, the Fermi energy in graphene is close to the Dirac point, enabling optical absorption due to interband transition. (c) The reflection intensity of the cavity at $\lambda$ = 1551 nm as a function of $V_G$.}
\end{figure}

We tested the temporal characteristics by observing the reflected intensity modulation of a narrowband tunable laser (1 MHz spectral bandwidth) coupled into the cavity. The cavity reflection was detected with an InGaAs avalanche photodetector (APD). To address the low coupling efficiency of the probe laser into the cavity, we improved the signal to noise ratio by locking our detection to a 20 kHz amplitude modulation of the probe laser power. To measure the low frequency ($<$3kHz) response, we combined a DC gate voltage of 5.7 V and a sinusoidal signal voltage of $V_{rms}$ = 0.1 V and then applied them across the graphene capacitor. Fig. 4(a) plots the normalized modulation depth for different input laser wavelengths, showing two peaks corresponding to the cavity resonances. For the measurement of high frequency response, we coupled a RF power of -10 dBm with varying DC gate voltages from a 20 GHz network analyzer (NWA) to the dual-layer graphene capacitor. The electrical signal of the InGaAs APD was sent to the input port of the NWA, the measurement result is shown in Fig. 4(b), indicating a high speed response of a low-pass filter characteristic with a 3 dB cut-off frequency of 1.2 GHz. The cut-off frequency is limited by the RC time constant of the dual-layer graphene capacitor, as deduced by the impedance measurement of the device (see supporting information).

\begin{figure}
  \includegraphics[width=17.0cm]{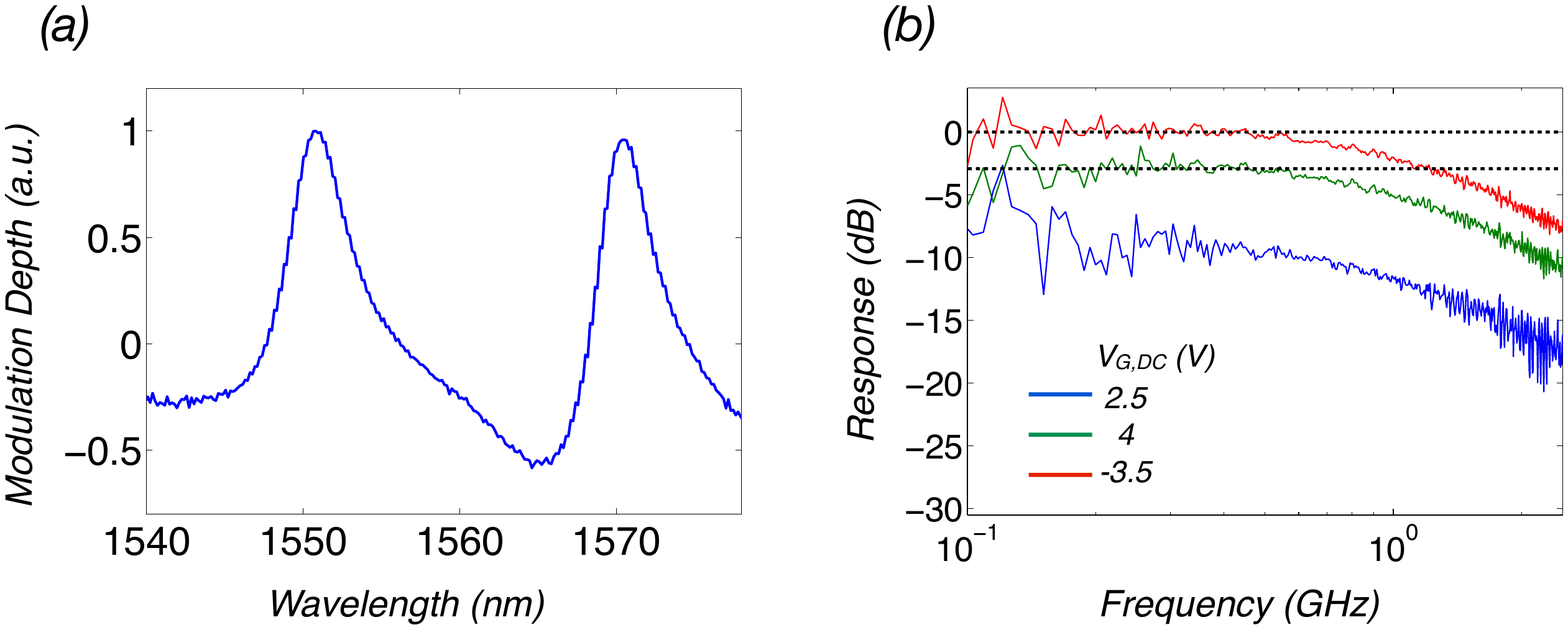}
  \caption{(a) Modulation depth as  function of input laser wavelengths measured at 3 kHz (b) High frequency response of the graphene modulator at different DC gate voltages. Dotted lines indicate the response of 0 dB and -3 dB.}
\end{figure}

In our device, the area of the graphene capacitor is about 100 $\upmu$m$^2$, corresponding to a capacitance of 320 fF. We estimate the switching energy of this device to be approximately 1 pJ (see supporting information). For PPC L3 cavities, the resonant mode area is around 0.5 $\upmu$m$^2$. Therefore, by shrinking the size of the graphene capacitor to only overlap with the mode area, we expect the switching energy to be only 5 fJ and operation spped at 100 GHz (see supporting information). The cavity bandwidth in this work exceeds 600 GHz for a $Q$ value of 300: i.e., it would be possible to obtain a large modulation contrast without the need of high $Q$ cavities, as is required in silicon carrier-depletion (injection) modulators. In addition, this broader bandwidth improves temperature stability, which is a limiting factor in the carrier-modulation Si modulators\cite{Manipatruni2008,Teng2009,Reed2010}.

In conclusion, we have demonstrated a high-speed, graphene-boron nitride heterostructure based electro-optic modulator integrated with a photonic crystal nanocavity with micrometer scale device footprint. The maximum modulation depth is 3.2 dB with a voltage swing of only 2.5 V. The modulation originates from the variation of optical conductivity of graphene under the effect of different electrostatic potential and the strong coupling between graphene and PPC cavity. At high frequencies, the device exhibits a 3 dB cut-off frequency up to 1.2 GHz. We estimate that by shrinking the device size from 100 $\upmu$m$^2$ to 0.5 $\upmu$m$^2$, the operating speed and switching energy could be improved by two orders of magnitude. These results show that the strong electro-absorptive effect of graphene in a cavity could create efficient optical modulation for on-chip modulators with ultra-low power consumption, ultra-small footprint, high-speed, and relatively broad bandwidth, which make such devices promising for efficient and stable electrical to optical signal conversion in optical communications and signal processing networks.


\acknowledgments{Financial support was provided by the Office of Naval Research (Award N00014-13-1-0662), Air Force Office of Scientific Research PECASE (supervised by G. Pomrenke), the DARPA Information in a Photon programme (grant no. W911NF-10-1-0416) and by NSF grant DMR-1106225 (T.H.). Device fabrication was partly carried out at the Center for Functional Nanomaterials, Brookhaven National Laboratory, which is supported by the US Department of Energy, Office of Basic Energy Sciences (contract no. DE-AC02- 98CH10886). Device assembly (including graphene transfer) and characterization was supported by the Center for Re-Defining Photovoltaic Efficiency Through Molecule Scale Control, an Energy Frontier Research Center funded by the US Department of Energy, Office of Science, Office of Basic Energy Sciences (award no. DE-SC0001085). R.-J.S. was supported in part by the Center for Excitonics, an Energy Frontier Research Center funded by the US Department of Energy, Office of Science, Office of Basic Energy Sciences under award no. DE-SC0001088.
}

\bibliography{library}


\end{document}